\documentclass[journal, twoside]{IEEEtran}

\usepackage{cite}
\usepackage{amsmath,amsthm,amssymb}
\usepackage{graphicx}\graphicspath{{figure/}{figure_case_study/}}
\usepackage{siunitx}
\usepackage{multirow}
\usepackage{stfloats}
\usepackage[para]{threeparttable}
\usepackage{cases}
\usepackage[thinlines,thiklines]{easybmat}
\usepackage{physics}
\usepackage{mathtools}
\usepackage{soul}
\usepackage{color}
\usepackage[table,xcdraw]{xcolor}
\usepackage{changes}
\definecolor{blueshade}{rgb}{1,0.8,0.8}
\sethlcolor{blueshade}

\newcommand{\bI}{\mathbf{I}}
\newcommand{\bJ}{\mathbf{J}}
\newcommand{\tbJ}{\tilde{\bJ}}
\newcommand{\bH}{\mathbf{H}}

\newcommand{\bM}{\mathbf{M}}

\newcommand{\bE}{\mathbf{E}}
\newcommand{\bF}{\mathbf{F}}

\newcommand{\by}{\mathbf{y}}
\newcommand{\bz}{\mathbf{z}}

\newcommand{\bg}{\mathbf{g}}
\newcommand{\bk}{\mathbf{k}}
\newcommand{\bl}{\mathbf{l}}

\newcommand{\bv}{\mathbf{v}}
\newcommand{\bw}{\mathbf{w}}

\usepackage[hidelinks]{hyperref}

\begin{document}
\title{Semi-implicit Continuous~Newton~Method for Power Flow Analysis}
\author{Ruizhi~Yu, Wei~Gu, Yijun~Xu, Shuai~Lu, Suhan~Zhang
    \thanks{This work is supported by the 2023 Graduate Research Innovation Plan of Jiangsu Province (Grant No. KYCX23\_0247).}
    \thanks{Ruizhi Yu, Wei Gu, Yijun Xu and Shuai Lu are with the school of electrical engineering, Southeast University, Nanjing, Jiangsu, 210096 China. \emph{Corresponding author: Wei Gu}, e-mail: wgu@seu.edu.cn.}
    \thanks{Suhan Zhang is with the faculty of engineering, The Hong Kong Polytechnic University.}
}
\maketitle
\begin{abstract}
    As an effective emulator of ill-conditioned power flow, continuous Newton methods (CNMs) have been extensively investigated using explicit and implicit numerical integration algorithms.
    Explicit CNMs are prone to non-convergence issues due to their limited stable region, while implicit CNMs introduce additional iteration-loops of nonlinear equations. 
    {Faced with this, we propose a semi-implicit version of CNM. We formulate the power flow equations as a set of differential algebraic equations (DAEs), and solve the DAEs with the stiffly accurate Rosenbrock type method (SARM). The proposed method succeeds the numerical robustness from the implicit CNM framework while prevents the iterative solution of nonlinear systems, hence revealing higher convergence speed and computation efficiency.
    A new 4-stage 3rd-order hyper-stable SARM, together with a 2nd-order embedded formula to control the step size, is constructed to further accelerate convergence by tuning the damping factor. 
    Case studies on ill-conditioned systems verified the alleged performance. An algorithm extension for MATPOWER is made available on Github for benchmarking.}
\end{abstract}
\begin{IEEEkeywords}
    Continuous Newton Method, Power flow Analysis
\end{IEEEkeywords}
\vspace{-0.8cm}
\section{Introduction}
\IEEEPARstart{P}{ower} flow lays the foundation for electric power system analysis\cite{Liuwanbin2024, Liuchengxi2024, Huangyujia2021}. {However, the Newton-Raphson method typically fails in ill-conditioned power flow cases. This is because, after formulating the power flow models as the solution of ordinary differential equations (ODEs)\mbox{\cite{federico2009}}, the Newton-Raphson iteration can be viewed as the explicit forward Euler numerical-integration step that suffers from poor stability. Therefore, non-convergence issues cannot be avoided.}

{From the numerical-integration point of view, we can enhance the convergence by substituting schemes with enlarged stability region for the forward Euler, which are called the continuous Newton methods (CNMs). Efforts are based on the explicit Runge-Kutta method (ERKM)\mbox{\cite{federico2009,TOSTADOVELIZ20202s4, TOSTADOVELIZ2019RKCOMP}}, the modified Euler method\mbox{\cite{TOSTADOVELIZ2019MODIFEULER}}, the midpoint rule\mbox{\cite{TOSTADOVELIZ2019RBS}}, the implicit backward Euler method (BEM)\mbox{\cite{federico2019}}, just to name a few. However, none of these above methods is both implicitly stable and explicitly solvable.}

{To overcome these issues, we propose a semi-implicit continuous Newton method (SICNM) based on a stiffly accurate Rosenbrock-type method. The proposed method requires no numerical iterations of nonlinear equations while it possesses a higher convergence speed and hyper-stability. The power flow results reveal its excellent performance.}
\section{{Preliminaries}}
{Let us first review the CNMs. Specifically, the power flow problem $\mathbf{0}=\bg(\by)$ with initial guess $\by_0$ is formulated as the equivalent initial value problem of ODE}
\begin{equation}
{
\dot{\by}=-\bJ^{-1}(\by)\bg(\by),\quad \by(0)=\by_0,
}
\label{eode}
\end{equation}
{where $\bJ$ is the Jacobian of $\bg$. The equilibrium state of \mbox{\eqref{eode}} is proved to be the solution to the original power flow\mbox{\cite{federico2009}}.}

{The explicit CNMs (ECNMs) convert \mbox{\eqref{eode}} into}
\begin{equation}
{
\by_{i+1}=\bF\qty(h,\by_i),
}
\label{ecnm}
\end{equation}
{where $\bF$ is the explicit scheme\mbox{\cite{federico2009,TOSTADOVELIZ20202s4, TOSTADOVELIZ2019RKCOMP,TOSTADOVELIZ2019MODIFEULER,TOSTADOVELIZ2019RBS}}, $h$ is the step size. To solve \mbox{\eqref{ecnm}} is explicitly recursive but the stability region of explicit scheme is limited, which means potential divergence. Also, we have to factorize the Jacobian several times in each step as $\bF$ contains $\bJ^{-1}$, which is computationally burdensome.}

{To mitigate the stability deficiency, the implicit CNM (ICNM) uses the hyper-stable BEM to solve the implicit ODEs }
\begin{equation}
{
\bJ(\by)\dot{\by}=-\bg(\by),
}
\label{iode}
\end{equation}
{where $\bJ^{-1}$ is eliminated by multiplying $\bJ$ to both sides of \mbox{\eqref{eode}}\mbox{\cite{federico2019}}. The derived scheme is}
\begin{equation}
{
\bJ(\by_{i+1})\qty(\by_{i+1}-\by_i)=-h\bg(\by_{i+1}).
}
\label{icnm}
\end{equation}
{That is, an inner iteration loop, which needs to factorize the Jacobian of $\bJ(\by)\by$, is required to solve for $\by_{i+1}$, adding extra divergence risks and being time-consuming. The dishonest factorization is utilized in\mbox{\cite{federico2019}} to relieve the burden but may impact the overall robustness.}

\vspace{-0.8cm}
\section{Semi-implicit Continuous Newton Method}
\vspace{-0.3cm}
\subsection{Conceptual Description}
\vspace{-0.2cm}
By introducing $\bz=\dot{\by}$, we rewrite \eqref{iode} as {the initial value problem of differential algebraic equations (DAEs)}
\begin{equation}
    \begin{cases}
        \dot{\by}=\bz\\
    0=\bJ(\by)\bz+\bg(\by)
    \end{cases},\label{dae}
\end{equation}
{where $\by(0)=\by_0$, $\bz(0)=\bz_0=-\bJ(\by_0)^{-1}\bg(\by_0)$.}

\subsubsection{The stiffly accurate Rosenbrock type method (SARM)}
We use the stiffly accurate Rosenbrock type method to solve \eqref{dae}, which has the general formulae\cite{wanner1996solving}
\begin{equation}
    {
    \begin{aligned}
        \qty(\bM-h\gamma \tbJ_0)
        \begin{bmatrix}
        \bk_i\\
        \bl_i
        \end{bmatrix}=&h
        \begin{bmatrix}
        \bw_i\\
        \bJ(\bv_i)\bw_i+\bg(\bv_i)
        \end{bmatrix}\\
        +&h\bJ_0
        \sum_{j=1}^{i-1}\gamma_{ij}
        \begin{bmatrix}
        \bk_j\\
        \bl_j
        \end{bmatrix},\ i=1,\dots, s,\\ 
        \begin{bmatrix}
            \bv_i\\
            \bw_i
        \end{bmatrix}
        =&
        \begin{bmatrix}
            \by_0\\
            \bz_0
        \end{bmatrix}
        +\sum_{j=1}^{i-1}\alpha_{ij}
        \begin{bmatrix}
        \bk_j\\
        \bl_j
        \end{bmatrix},\\
        \begin{bmatrix}
            \by_1\\
            \bz_1
        \end{bmatrix}
        =&
        \begin{bmatrix}
            \by_0\\
            \bz_0
        \end{bmatrix}
        +\sum_{i=1}^{s}b_i
        \begin{bmatrix}
        \bk_i\\
        \bl_i
        \end{bmatrix},\\
    \end{aligned}
    }\label{sarm1}
\end{equation}
where $s$ is the stage number; $h$ is the step size; $\tbJ_0$ is the Jacobian of \eqref{dae} evaluated at $(\by_0, \bz_0)$ with
\begin{equation}
    {
    \tbJ_0=
    \begin{bmatrix}
        \mathbf{0}&\mathbf{I}\\
        \bH(\by_0)\otimes\bz_0+\bJ(\by_0)&\bJ(\by_0)
    \end{bmatrix};}\label{J0}
\end{equation}
\[
{
    \bM=
    \begin{bmatrix}
    \bI&\mathbf{0}\\
    \mathbf{0}&\mathbf{0}
    \end{bmatrix};}
\]
$\otimes$ is the Kronecker product; $\bH\qty(\by)$ is the Hessian tensor of power flow with
\[
    \bH\qty(\by)=\qty(\pdv{\grad \bg_i(\by)^\mathrm{T}}{\by_j})_{ij},
\]
the analytical derivation of which can be found in \cite{src_codes};
{$\bI$ is the identity matrix;
$\alpha_{ij}$, $\gamma_{ij}$, $\gamma$, $b_i$ are coefficients controlling the accuracy and numerical stability of the method.}

The method reads \emph{semi-implicit} because it possesses the exclusive hyper-stability, which will be shown later, of implicit method, {whilst the stage-by-stage calculations of $\bk_i$ and ${\bl}_i$ in \mbox{\eqref{sarm1}} are explicit recursions since they depend at most on $\bk_{i-1}$ and ${\bl}_{i-1}$.}

\subsubsection{Partial factorization}
The matrix $\tilde{\bE}=\qty(\bM-h\gamma \tbJ_0)$ is constant in each step. 

Denoting the $i$-$j$ block of \eqref{J0} by $\tbJ_{ij}$ and performing the block LDU factorization, we have
\[
    \tilde{\bE}=
    \begin{bmatrix}
        \bI&\mathbf{0}\\
        -h\gamma \tbJ_{21}&\bI
    \end{bmatrix}
    \begin{bmatrix}
        \bI&\mathbf{0}\\
        \mathbf{0}& -h\gamma\qty(h\gamma\tbJ_{21}+\tbJ_{22})
    \end{bmatrix}
    \begin{bmatrix}
        \bI&-h\gamma\bI\\
        \mathbf{0} &\bI
    \end{bmatrix}.
\]
That is, only one LU factorization of matrix $-h\gamma\qty(h\gamma\tbJ_{21}+\tbJ_{22})$, which has the same size of system \eqref{iode}, is required in each step. Hence, the problem scales remain unchanged by transformation \eqref{dae}.

\subsubsection{Adaptive Step Size by Embedded Solutions}
Suppose the order of \mbox{\eqref{sarm1}} is $q$, we obtain $\by_1$ and $\bz_1$ with error $\order{h^{q+1}}$.
By choosing proper $\alpha_{si}$, we can construct $\hat{\by}_1=\by_0+h\sum_{i=1}^{s-1}\alpha_{si}{\bk}_i$ and $\hat{\bz}_1=\bz_0+h\sum_{i=1}^{s-1}\alpha_{si}{\bl}_i$ with error $\order{h^{q}}$. Then the difference $\Delta \by=|\hat{\by}_1-\by_1|$ and $\Delta \bz=|\hat{\bz}_1-\bz_1|$ can be used as practical estimations of $\order{h^{q}}$ to adjust the step sizes with formula

\[h_\text{new}=h\cdot \qty(\qty|\frac{[\Delta \by\ \Delta \bz]^\text{T}}{Atol+Rtol\cdot |[\by_1\ \bz_1]^\text{T}|}|_\infty)^{-\frac{1}{q}},\]
where $Atol$ and $Rtol$ are respectively the absolute and relative error tolerance, which requires no extra computation burden. 

\subsection{Construction of Rodas3d}
We seek for a SARM with 3rd-order because SARMs with higher order require at least two more stages\cite{wanner1996solving,GerdSteinebach2020,GerdSteinebach2023}. Also, we do not care about the $\by$, $\bz$ trajectories 
but only the speed to reach the equilibrium state, which means the modals of \eqref{dae} should be damped as soon as possible. Therefore, we tune the parameter $\gamma$ and construct a 4-stage 3rd-order 2nd-embedded-order SARM, the Rodas3d, for better damping performance in the context of power flow analysis. 

Following \cite{wanner1996solving}, we have the following order conditions for the desired Rodas3d
\[
    \begin{cases}
        b_1+b_2+b_3+b_4=1\\
        b_2\beta_2'+b_3\beta_3'+b_4\beta_4'=1/2-\gamma\\
        b_2\alpha_2^2+b_3\alpha_3^2+b_4\alpha_4^2=1/3\\
        b_3\beta_{32}\beta_2'+b_4\beta_{43}\beta_3'+b_4\beta_{42}\beta_2'=1/6-\gamma+\gamma^2\\
        \sum_{i=1}^3\alpha_{4i}=1,\quad \alpha_{31}+\alpha_{32}=1\\
        \alpha_{42}\beta_2'+\alpha_{43}\beta_3'=1/2-\gamma\\
        b_{i}=\beta_{4i}\quad i=1,2,3,4,\quad\alpha_{4i}=\beta_{3i}\quad i=1,2,3\\
    \end{cases}
\]
where $\alpha_i=\sum\nolimits_{j=1}^{i-1}\alpha_{ij}$, $\beta_{ij}=\alpha_{ij}+\gamma_{ij}$ and $\beta_i'=\sum_j\beta_{ij}$.

Rodas3d has the stability function
\[R(z)=\frac{\splitfrac{1+(1-4\gamma)z+(6\gamma^2-4\gamma+1/2)z^2}{+(-4\gamma^3+6\gamma^2-2\gamma+1/6)z^3}}{(\gamma z-1)^4}.\]
It is hyper-stable by nature because $R(\pm\infty)=0$, which means it damps both stable and unstable modals.
And we can confine the damping region more precisely so that all the stable modals are contained, which, according to Table 6.4 of \cite{wanner1996solving}, requires $0.22364780\leq \gamma\leq 0.57281606$. 
We choose $\gamma=0.57281606$ so that Rodas3d can more obviously damp the modals with large eigenvalue modules, which is verified to be effective by trials. One of the solutions to the above order condition equations can be found in our implementation\cite{src_codes}.

\begin{figure}[!h]
    \centering
    \includegraphics[width=3.4286in]{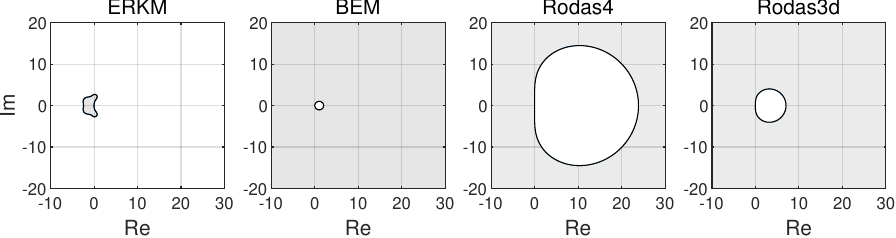}
    \caption{{The stability region indicated by shading.}}
    \label{sr}
\end{figure}

{The stability region of different methods are shaded in Fig.\mbox{\ref{sr}}, where Rodas4 is the famous representative of SARM with 6-stage and 4th-order\mbox{\cite{wanner1996solving}}. The points in the shaded area are modals that can be damped by the method. Since we want to damp both stable and unstable modals as explained above, the shaded area should be as big as possible. Obviously, the stability region of the ERKM is quite limited due to its explicit nature. By adjusting the damping factor, Rodas3d reduces the unstable region, bringing it closer to that of the BEM and outperforming the Rodas4.}

\begin{table*}[]
\begin{threeparttable}
\centering
\caption{Iteration Times and Total Computation Overhead}
\label{Comp_ite}
\begin{tabular}{ccccccccccccc}
\hline
\multirow{2}{*}{Case} & \multicolumn{1}{c}{\multirow{2}{*}{M1}} & \multicolumn{1}{c}{\multirow{2}{*}{M2}} & \multicolumn{1}{c}{\multirow{2}{*}{M3}} & \multicolumn{1}{c}{\multirow{2}{*}{M4}} & \multicolumn{1}{c}{\multirow{2}{*}{M5}} & \multicolumn{1}{c}{\multirow{2}{*}{M6}} & \multicolumn{4}{c}{M7}                                                                                               & \multicolumn{2}{c}{M8}                                   \\ \cline{8-13} 
                      & \multicolumn{1}{c}{}                    & \multicolumn{1}{c}{}                    & \multicolumn{1}{c}{}                    & \multicolumn{1}{c}{}                    & \multicolumn{1}{c}{}                    & \multicolumn{1}{c}{}                    & \multicolumn{1}{c}{ICNM-JH} & \multicolumn{1}{c}{ICNM-J} & \multicolumn{1}{c}{ICNM-J1} & \multicolumn{1}{c}{ICNM-J0} & \multicolumn{1}{c}{Rodas4} & \multicolumn{1}{c}{Rodas3d} \\ \hline
3012wp                & D.\tnote{a}                                    & NC.\tnote{b}                               & D.                                    & D.                                    & D.                                    & NC.                               & 79(6.16s)                   & 178(11.34s)                & NC.                 & NC.                   & NC.                  & 23(1.19s)                   \\
9241pegase            & D.                                    & NC.                               & D.                                    & D.                                    & D.                                    & NC.                             & 91(23.66s)                  & 189(39.91s)                & NC.                 & NC.                 & 25(4.49s)                  & 19(2.95s)                   \\
ACTIVSg25k            & D.                                    & NC.                               & D.                                    & D.                                    & D.                                    & NC.                               & 97(72.50s)                  & 224(140.77s)               & NC.                 & NC.                   & 27(15.65s)                 & 19(9.36s)                   \\ \hline
\end{tabular}
\smallskip
\scriptsize
\begin{tablenotes}
\item[a] Divergent
\item[b] Not convergent after 1000 iterations
\end{tablenotes}
\end{threeparttable}
\end{table*}

\begin{table*}[]
\centering
 \caption{Rates of Convergence In Limit Tests}
\begin{tabular}{cccccccccccc}
\hline
\multirow{2}{*}{M1} & \multirow{2}{*}{M2} & \multirow{2}{*}{M3} & \multirow{2}{*}{M4} & \multirow{2}{*}{M5} & \multirow{2}{*}{M6} & \multicolumn{4}{c}{M7}               & \multicolumn{2}{c}{M8} \\ \cline{7-12} 
                    &                     &                     &                     &                     &                     & ICNM-JH & ICNM-J & ICNM-J1 & ICNM-J0 & Rodas4    & Rodas3d    \\ \hline
0\%                 & 12.1\%                & 22.2\%                & 0.3\%                 & 0.9\%                 & 0.6\%                 & 3.1\%     & 0\%    & 0\%     & 0\%     & 53.8\%      & 59.4\%       \\ \hline
\end{tabular}
\label{Conv_rate}
\end{table*}
\section{Case Studies}
To validate the performance of our proposed method, we implemented the following classic and state-of-the-art numerical methods for comparison:
\begin{itemize}
    \item M1: the standard Newton-Raphson method;
    \item M2: the Iwamoto's robust method \cite{Iwamoto1981}; 
    \item M3: the ECNM based on ERKM \cite{federico2009};
    \item M4: the Romberg-integration-based method \cite{TOSTADOVELIZ2020Romberg};
    \item M5: the modified CNM based on the second-stage fourth-order explicit Runge-Kutta \cite{TOSTADOVELIZ20202s4};
    \item M6: the Mann-Iteration-based method \cite{TOSTADOVELIZ2021mann};
    \item M7: the ICNM series with adaptive step size\cite{federico2019};
    \item M8: Our proposed SICNM series based on Rodas4\cite{wanner1996solving} and our constructed Rodas3d.
\end{itemize}
We implemented all the methods in MATPOWER V7.1 \cite{matpowerACpowerflow}. For M2 to M7, we used the parameter setting from the original paper. We found that to set initial step size $1$ for M7, as suggested in \cite{federico2019}, was too big to maintain convergence. 
As a result, we set the initial step size of M7 to be $0.01$ since it was adjusted adaptively therein. The $Atol$ and $Rtol$ of M8 were all $0.1$. 

{The reactive power limits were checked by MATPOWER routine. In the event of limit violation, the bus type was first converted and then the power flow was rerun.}
We performed all the tests on a desktop equipped with AMD Ryzen 3700X and 32 GB RAM. The coding environment was MATLAB R2024a. We set the error tolerance $\varepsilon=1e-5$. The codes and case data are made as the Github repository \text{MATPOWER-SICNM} for result reproducibility and benchmarking\cite{src_codes}.
\subsection{Ill-conditioned Bench Tests}
The first test was to solve three ill-conditioned benches based on the MATPOWER built-in data case \texttt{3012wp}, \texttt{9241pegase}\cite{Fliscounakis2013,josz2016ac} and \texttt{ACTIVSg25k}\cite{Case_ACTIVSg25k}. The simulation results, profiling the number of iterations and time costs required to converge, were shown in TABLE \ref{Comp_ite}. 

\begin{figure}[!h]
    \centering
    \includegraphics[width=3in]{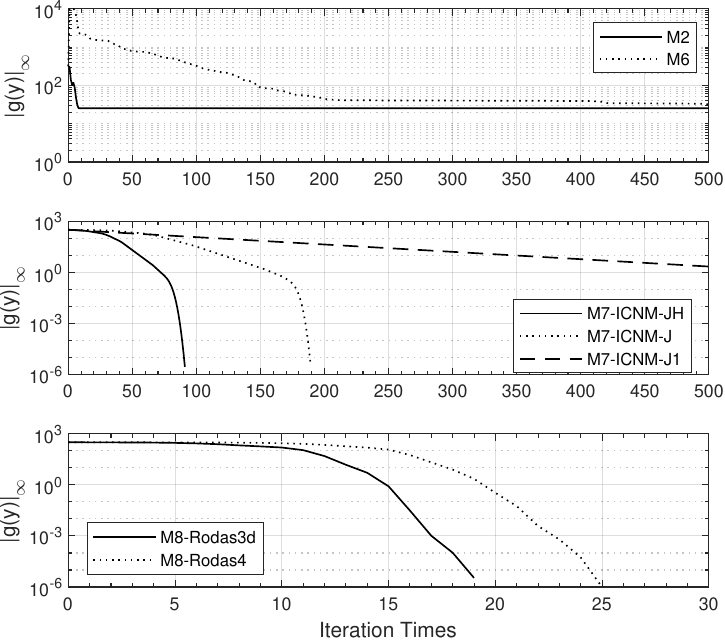}
    \caption{Convergence errors versus iteration times in Case 9241pegase.}
    \label{Converrors}
\end{figure}

M1 to M6 failed in all of the power flow simulations. It is illustrated in Fig. \ref{Converrors} that the error norms of M2 and M6 decreased fast in the early stage but soon stalled. The CNM-based methods with the implicitness peculiarity, M7 and M8, were more robust. 
Though slowly at the beginning, they sped up the convergence as iteration number increased.
It is conspicuous that M8 converged much faster than M7 and the 
Rodas3d-based one was the fastest. This is because: 1) first, Rodas4 and Rodas3d have more than 3rd-order accuracy which means it can predict the future state more precisely whereas the BEM is 1st-order only; 2) second, the damping factor of Rodas3d was carefully chosen in the construction stage; 3) third, the convergence of ICNM series cannot be theoretically guaranteed.

\begin{figure}[!h]
    \centering
    \includegraphics[width=3in]{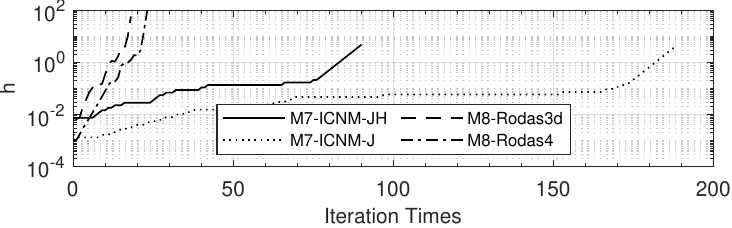}
    \caption{{Step size versus iteration times in Case 9241pegase.}}
    \label{h}
\end{figure}

The convergence speed can be further explained by Fig. \ref{h}. It is apparent that M8 used bigger average step sizes, which, however, was of magnitude 1e-2 at the very beginning. This should be attributed to the embedded solutions which estimated the local errors in an accurate but computationally cheap way so that the step sizes can be altered non-conservatively but within the stable region. In contrast, the strategy for BEM is empirical and conservative. Also, we found that the choice of proper initial step size is fundamental in M7 implementation, because big initial step sizes would lead to divergence while small step sizes may reduce the convergence speed. The fine tuning of M7 relies on massive trial-and-errors from case to case, 
while M8 {can choose proper step size by itself}.

Moreover, we found in Table \ref{Comp_ite} that M8 shows apparent efficiency advantages over M7. {This can be explained by TABLE \mbox{\ref{stats}}. Apparently, only one LU factorization of the matrix with the original system size was performed in each step of M8 and the total iteration times were greatly reduced. The Rodas3d-based M8 was the most efficient because it has fewer computation stages, greatly reducing the function evaluations.}
\begin{table}[!h]
\centering
\caption{{Number of Function Evaluations in Case 9241pegase}}
\begin{tabular}{lcccccc}
\hline
           & \multirow{2}{*}{$\bg$} & \multirow{2}{*}{$\bJ$} & \multirow{2}{*}{$\bH(\by)\otimes \bz$} & LU    & Reject             & Accept \\
           &                        &                        &                                        & Fact. & Steps              & Steps  \\ \hline
M7-ICNM-JH & 330                    & 330                    & 330                                    & 330   & N\textbackslash{}A & 91     \\
M7-ICNM-J  & 742                    & 742                    & 0                                      & 742   & N\textbackslash{}A & 189    \\
M8-Rodas4  & 222                    & 247                    & 25                                     & 37    & 12                 & 25     \\
M8-Rodas3d & 124                    & 143                    & 19                                     & 31    & 12                 & 19     \\ \hline
\end{tabular}
\label{stats}
\end{table}

\subsection{Limit Tests with Random Initial Feeding}
Let us verify the limit performance of our proposed method. We perturbed the initial values of case \texttt{9241pegase} by adding deviations obeying normal distribution within $[-0.005,0.005]$ rad to angles of randomly-picked 1/2 non-slack buses. A thousand times of simulations with such initial value settings were performed with error tolerance 1e-5. The initial step sizes of all methods were set to be $0.1$ if applicable. The rate of convergence within 40 iterations were profiled in TABLE \ref{Conv_rate}. The proposed M8 achieved the highest rate of convergence. The Rodas3d-based M8 averagely converged faster than the Rodas4-based one in successful cases, as indicated by TABLE \ref{ave_ite}.

\begin{table}[!h]
\centering
\caption{Iteration Times and Total Computation Overhead Averaged in Limit Tests}
\label{ave_ite}
\begin{tabular}{cc}
\hline
\multicolumn{1}{c}{Rodas4-based M8} & \multicolumn{1}{c}{Rodas3d-based M8} \\ \hline
27.5(4.58s)                         & 22.0(3.11s)                          \\ \hline
\end{tabular}
\end{table}

\section{Conclusion}
In this paper, we propose a series of semi-implicit continuous Newton methods for ill-conditioned power flow analysis. Simulations revealed their superior capability in terms of convergence speed and robustness. The proposed method can serve as a potent backup to do cross-validation in the event of Newton-Raphson's failure. The MATPOWER-based implementation of the proposed method has been made public for whom may concern.

\section*{Acknowledgement}
The authors would like to thank Prof. Dr Gerd Steinebach from Hochschule Bonn-Rhein-Sieg, Sankt Augustin, Germany, for his providing paradigm of high-performance codes and theoretical support.
\bibliographystyle{IEEEtran}
\bibliography{IEEEabrv,bibliography}
\end{document}